\documentclass[runningheads]{llncs}
\usepackage{graphicx}
\pdfoutput=1

\usepackage{booktabs}
\usepackage{amsfonts}
\usepackage{times}
\usepackage{tikz} \usetikzlibrary{shapes,arrows,shapes.geometric,fit,calc,positioning,automata,shadows}
\usepackage{stackengine}
\usepackage{enumitem}
\usepackage{balance}
    \usepackage[misc,geometry]{ifsym} 
\usepackage{verbatim}
\usepackage{enumitem}
\usepackage{todonotes}
\usepackage{quoting,xparse}

\newtheorem{observation}{Observation}
\newcommand{\ISS}{\mathsf{ISS}}
\begin{document}
\title{An Introduction to Engineering  Multiagent  Industrial Symbiosis Systems: Potentials and Challenges }
\titlerunning{An Introduction to Engineering  Multiagent ISS: Potentials and Challenges}

\author{
Vahid Yazdanpanah(\Letter)\inst{1}\and Devrim Murat Yazan\inst{1} \and \\ Jos van Hillegersberg\inst{1}\and Mehdi Dastani\inst{2}}
\authorrunning{V. Yazdanpanah et al.}
\institute{University of Twente, Enschede, The Netherlands \\ 
\email{\{V.Yazdanpanah,D.M.Yazan,J.vanHillegersberg\}@utwente.nl} \and 
Utrecht University, Utrecht, The Netherlands \\
\email{M.M.Dastani@uu.nl}
}
\maketitle              % typeset the header of the contribution
\begin{abstract}

Multiagent Systems (MAS) research reached a maturity to be confidently applied to real-life complex problems. Successful application of MAS methods for behavior modeling, strategic reasoning,  and decentralized governance, encouraged us to focus on applicability of MAS techniques in a class of industrial systems and to elaborate on potentials and challenges for method integration/contextualization. We direct attention towards a form of industrial practices called \emph{Industrial Symbiosis Systems} ($\ISS$) as a highly dynamic domain of application for MAS techniques. In $\ISS$, firms aim to reduce their material and energy footprint by circulating reusable resources among the members. To enable systematic reasoning about $\ISS$ behavior and support firms' (as well as $\ISS$ designers') decisions, we see the opportunity for marrying \emph{industrial engineering} with \emph{engineering multiagent systems}. This enables introducing (1) representation frameworks to reason about dynamics of $\ISS$, (2) operational semantics to develop computational models for  $\ISS$, and (3) coordination mechanisms to enforce desirable $\ISS$ behaviors. We argue for applicability and expressiveness of resource-bounded formalisms and norm-aware  mechanisms for the design and deployment of $\ISS$ practices. In this proposal, we elaborate on different dimensions of $\ISS$, present a methodological foundation for $\ISS$ development, and finally discuss open problems.

\end{abstract}
%
%
%===== BEGIN : MAIN BODY ========%

%===== main body of paper ===============

%=============== INTRO ==================
\section{Introduction}

As a response to recent environmental concerns on reducing the exploitation and discharge of (pure) resources, the idea to circulate reusable resources (also known as the practice of circular economy \cite{kirchherr2017conceptualizing}) gained attention in both academia as well as industry. Industrial Symbiosis Systems ($\ISS$) are green/waste-free implementations of the concept of circular economy in the context of industrial relations \cite{chertow2007uncovering,yazan2016design}.  For instance, a cluster of firms $A$, $B$, and $C$ may have the chance to reuse each other's wastes. E.g., $A$'s  excess steam can be used in $B$'s turbines, $B$'s waste water can be used for cooling  in  $C$, and $C$'s sludge can substitute a primary input of $A$. Exploiting such an opportunity may lead to financial as well as socio-ecological gains.   As $\ISS$ is merely focused on circulating non-commoditized resources, e.g., waste material and energy,  waiting for traditional market structures to materialize and successfully implement $\ISS$ would be a guaranteed failure\footnote{In a more concrete sense,  given an environmentally desirable collaboration among $A$, $B$, and $C$, the allocation of collectively obtained benefits could be problematic in case any firm (group) can gain more by defecting the $\ISS$. In the game-theoretical language, given an empty core of the coalitional game, guaranteeing the stability of the collaboration requires external incentives (see \cite{DBLP:conf/atal/YazdanpanahYZ18}), hence calls for  contextualized methods for incentive engineering in $\ISS$.}. Thus, $\ISS$ have to be well-designed to be sustainable and in some cases its stability requires the introduction of external monetary incentives \cite{DBLP:conf/atal/YazdanpanahYZ18}. As Eric Maskin says: ``\textit{The market is no god---it cannot solve every problem}". This contribution calls for engineering  market places for $\ISS$ to address four main dimensions of such systems \cite{chertow2007uncovering}: temporal evolution, strategic aspects, resource boundedness, and normativity. 

%==========aspects==================

\noindent {\textbf{Temporal Evolution (TE):}} Since $\ISS$ builds upon potentially long-term industrial synergies, it is not similar to single-shot industrial transactions but has a temporal dimension. The preference on establishing long-term relations is mainly due to presence of additional costs (such as transportation, treatment, and transaction costs) for  implementing  $\ISS$. Roughly speaking, firms prefer to implement a relation that lasts---aiming to avoid renegotiating new relations for each and every transfer of reusable resources. Moreover, $\ISS$ operates  within the temporally dynamic industrial context (with respect to costs, demands, and supply). Thus, engineering attempts to ensure $\ISS$ strength, stability, and fairness have to take into account the dynamicity of such systems over time. 

\noindent {\textbf{Strategic Aspects (SA):}} The initiation and stability of relations in $\ISS$ critically depends on  the dynamics of power and firms' strategic abilities. In principle, firms act with respect to their preferences and strategic potentials---which in turn defines their potential for success\footnote{See \cite{miller1981power} for the interrelations between preferences,  power, and success.} and their so-called \emph{strategic responsibility} \cite{strategic_resp}. Dismissing this aspect leads to (agent-based) models that represent all firms in the same (power) level and abstract from the reality of industrial organizations. This call for game forms able to capture the dynamics of power, executable plans/strategies, and  firms' potential outcomes (i.e., to  specify and analyze ``what strategy ensures what situation for what group of firms''). 

\noindent {\textbf{Resource Boundedness (RB):}} In $\ISS$, the implementability of relations  is bounded to \emph{physical} (i.e., to be circulated) as well as \emph{financial} (i.e., to be obtained/invested) resources. Basically, firms have to take into account whether their short-term decisions/actions and long-term plans/strategies are in-line with their available resources. To capture this, a natural approach would be to model $\ISS$ using modeling structures rooted in resource-bounded logics \cite{alechina2010resource}.

\noindent {\textbf{Normativity and Coordination (NC):}} In brief,  normativity of $\ISS$ refers to the argument that: ``\emph{not all forms and potential implementations of industrial symbiosis are environmentally/socio-economically desirable}''. In Section \ref{sec:normativuty}, we elaborate on this argument and motivate the need for tailoring coordination mechanisms to nudge the behavior of $\ISS$ towards collectively desirable ones\footnote{As we elaborate later, resource-bounded formalisms capture TE, SA, and RB but we lack  frameworks able to integrate  NC as well.}. 

The literature on industrial symbiosis is rich in addressing case-specific issues. However, we see that focusing on implementing specific cases results in hardly  generalizable solution concepts. E.g., focusing on single-shot 1-1 relations dismisses the temporal as well as strategic (coalition-formation) dynamics in $\ISS$. We observe the lack of  generic methods---able to capture the above mentioned dimensions (i.e., TE, SA, RB, and NC)---for  designing and coordinating $\ISS$. In principle, $\ISS$ is a highly dynamic MAS, as it demonstrates normative and strategic dynamics in a temporal and resource-bounded context. 

To address these dimensions, we follow the idea that the formal representation of multi-stakeholder systems \cite{ferber2003agents,horling2004survey} enables systematic reasoning about their evolution  and fosters coordinating  their complex behavior. This motivates the need for introducing a new line of scientific inquiry on \textit{multiagent models for industrial symbiosis systems}. In the following, we elaborate on main requirements for engineering  such systems and  propose a methodological foundation for it. We also discuss that this domain of multiagent systems application calls for integrating some (not yet integrated) multiagent formalisms.

\section{A Normative Multiagent Approach}

We present a methodological foundation to enable modeling $\ISS$ and specifying its temporal, resource-bounded, and strategic aspects. Building on this, we discuss the normativity of industrial symbiosis and motivate the need for employing \emph{norm-aware coordination} mechanisms.

\subsection{Modeling ISS} 

We see $\ISS$ as a multiagent system in which each agent's action ``consumes'' resources---either financially, physically, or both. A rich method for representing and reasoning about such systems is Resource-Bounded Alternating-time Temporal Logic (RB-ATL)---introduced in \cite{alechina2010resource}. Building on an adopted version of the semantic machinery of RB-ATL, an $\ISS$ can be modeled as a Resource-Bounded Concurrent Game Structure (RB--CGS\cite{alechina2010resource}) $\mathcal{M}=(N,Res,Q,Act,d,c,o)$ where: $N=\{a_1,$ ... $,a_n\}$ is the set of $n$ firms involved in $\ISS$; $Res=\{res_1,\dots,res_r\}$ is a set of $r$ resource types; $Q$ is a non-empty set of states; $Act$ is a non-empty set of actions; $d: Q \times N \mapsto \mathcal{P}(Act) \setminus \{\emptyset\}$ is a function that in each state $q \in Q$, assigns a non-empty set of actions $d(q,a) \subseteq Act$ to each firm $a \in N$ (by $D(q)$, we denote the set of all joint actions $\sigma=(\alpha_1,\dots,\alpha_n)$ such that $\alpha_i \in d(q,a_i)$); partial function $c : Q \times N \times Act \mapsto \mathbb{N}^r$  maps a state $q \in Q$, firm $a \in N$, and an action $\alpha \in d(q,a)$ to a vector of integers, where a positive integer in position $i$ indicates the consumption of $res_i \in Res$ by action $a$; and $o : Q \times D(q) \mapsto Q$ is a deterministic transition function---modeling the evolution of the $\ISS$---that for each state  $q \in Q$ and joint action $\sigma \in D(q)$, returns a state that results from the execution of $\sigma$ in $q$.

In our suggested formalization, $Res$ may consist of both financial as well as physical resources. Specifying such a set of consumable resources, modeling their consumption using the cost function $c$, and combining these two with ATL-based transition systems enables modeling the real-life behavior  of $\ISS$ as \emph{industrial organizations} \cite{tirole1988theory}. Such a  modeling framework is flexible with respect to the set of available actions ($Act$), that the $\ISS$ designer intends to consider in its model. For instance, one may limit $Act$ to physical actions (e.g., sending or recycling some quantity of a resource), consider communicative actions \cite{dignum1996modelling} (e.g., informing another agent and updating its belief-/knowledge-base), or specify a more complex mix of institutional and physical acts \cite{dastani2017commitments} (e.g., offering/accepting a deal or sending/receiving a resource). Note that the use of RB--CGS enables relying on computationally attractive algorithms for verifying temporal, resource-bounded,  and strategic properties  \cite{alechina2015symbolic}. Following the successful use of RB--CGS and related resource-aware action models to specify real-world systems \cite{alechina2019unbounded}, we propose contextualizing RB--CGS for $\ISS$ modeling and its integration with norm-aware  techniques for $\ISS$ coordination.

%===============================================%

\subsection{Normativity of ISS} \label{sec:normativuty}

Using this representation, one can reason about the evolution of an $\ISS$ (temporarily) in which firms are able to execute their decisional plans (strategic) that are bounded to consumable resources  (resource-boundedness). We argue that the autonomy of firms, the availability of resources, and the  evolution  of $\ISS$ raise the concern that we may see undesirable situations.

In principle---from the perspective of the $\ISS$ designer/engineer---an $\ISS$ may go to an (un)desirable  \emph{state}, a firm may execute an (un)desirable \emph{action}, or in a more complex level, an $\ISS$ may perform  an (un)desirable \emph{behavior}. This is, to ascribe \emph{normative} labels to states, actions, and behaviors.

\noindent {\textbf{State Normativity:}} In an $\ISS$, the modeler can represent potential implementations of a relation in states. For instance,   firms may either implement a fair  benefit allocation (represented by state $q_{f}$) or  otherwise occurs (represented by state $q_{u}$). Then, the system designer can aim for suppressing unfair relations by ascribing undesirability to $q_u$ and accordingly implement normative mechanisms to coordinate the behavior of $\ISS$.  

\noindent {\textbf{Action Normativity:}} In addition to the above-mentioned state normativity, the execution of some actions (in some specific states) can also be bad/good. For example, the act of transporting a hazardous resource or defecting from an unfulfilled commitment (i.e., when the promise is not yet satisfied)  can both be seen undesirable. More generally, a subset of available actions (to an agent in a state) can be labeled as undesirable.  

\noindent {\textbf{Behavior Normativity:}}  Specifying state and action normativity is not sufficient to express some more complex undesirabilities (e.g., those that are concerned with the order of execution of some actions). In principle, the semantic notion of system's normative behavior is about the desirability of a chain of actions/events/states.  For instance, even if neither \emph{sending} an  offer, \emph{receiving}  an offer, nor \emph{accepting}  the offer are bad actions, their execution in a wrong order can be an undesirable system behavior (e.g., the chain of events in which a firm first accepts an offer and then receives it can be an undesirable one). 
%
%\end{itemize}
%
Note that $\ISS$ is designed to bring about a stable and resilient practice of circular economy, hence has to be aware of---and able to control---such normative situations.

\subsection{ISS Behavior and Normative Semantics}

First, we show how our suggested modeling structure enables specifying the $\ISS$ behavior as well as its normativity. 

\begin{definition}[$\ISS$ Behavior] Given an $\ISS$ $\mathcal{M}$, we denote the set of infinite sequences of states by $Q^\omega$. A sequence $\lambda=q_0,q_1,\dots \in Q^\omega$ is an \emph{$\ISS$ behavior} iff for all tuples of states $(q_i, q_{i+1})$, there exist a $\sigma \in D(q_i)$ such that $o(q_i,\sigma)=q_{i+1}$. The set of all possible behaviors of $\mathcal{M}$ is denoted by $\Lambda_{\mathcal{M}}$ ($\Lambda$ if $\mathcal{M}$ is clear from the context).
\end{definition}

While in some cases, specifying the normativity of states/actions is sufficient, the semantically rich notion of behavior normativity enables reasoning about and coordinating the more complex behavioral aspect of $\ISS$. 

\begin{definition}[$\ISS$ Behavioral Norm, Compliance, Violation] A norm $\mathcal{N}$ is a subset of $\ISS$ behaviors $\Lambda$ (i.e., $\mathcal{N} \subseteq \Lambda$). An $\ISS$ behavior $\lambda$ complies with $\mathcal{N}$ (for short, $\lambda$ is $\mathcal{N}$-compliant) iff $\lambda \in \mathcal{N}$. Otherwise, $\lambda$  violates $\mathcal{N}$ (for short, $\lambda$ is $\mathcal{N}$-violation).
\end{definition}

Although the subject  of norm synthesis/evolution in $\ISS$ is important, in this proposal we assume norms to be given and fixed. Such norms can be either derived using data-driven methods or defined from the perspective of a  third-party/legislative agent\footnote{One may construct a norm considering the consumption of resources  (e.g., if a behavior is in-line with available resources in $Res$), may model a norm to specify performable  behaviors  under a budget limit (e.g., if aggregating all the costs---modeled by the $c$ function---is under a threshold), or may combine the two  to formulate  more complex  norms.}. 

\begin{observation}[Expressivity]
Specifying $\ISS$ behaviors (using RB--CGS) is also expressive for modeling state and action normativity. State normativity can be derived using a single-state $\ISS$ behavior while action normativity can be modeled based on a two-state $\ISS$ behavior and the action profile that connects the two.    
\end{observation}

Given a formal representation of $\ISS$, one may ask whether verifying normative propositions---e.g., the compliance of the $\ISS$ behavior to a given norm---are computationally feasible. Relying on well-established results of \cite{alechina2010resource}, we have that  Given an $\ISS$ $\mathcal{M}$, verifying the validity of (normative) propositions has the computational complexity of $O(\ell^{2|Res| +1} \times |\mathcal{M}|)$ where $\ell$ is the size of the formula.

We presented a methodological foundation, rooted in the semantic machinery of RB--CGS, for representing and reasoning about $\ISS$ behavior. Next, we propose that for maintaining the $\ISS$ behavior---within an acceptable spectrum---norm-based coordination mechanisms are applicable and appropriate system engineering tools.

%=====================================

\subsection{Coordinating ISS} \label{sec:coordinating}

While normativity refers to the property of $\ISS$ (behavior) to be desirable or undesirable, normative coordination mechanisms are mainly methods for flexible administration to keep the system (behavior) within an acceptable spectrum. Operationalizing norms and employing norm-based instruments to ensure desirable system behaviors is a well-established principle in multiagent systems research. This includes proposals on applicability of normative coordination methods as a tool for operating electronic institutions \cite{aldewereld2007operationalisation}, for monitoring commitments in multiagent organizations \cite{dastani2017commitments}, and for implementing protocols in  norm-governed networks \cite{artikis2004protocol,yazdanpanah2016normative}.

Relying on this literature and considering the normative characteristics of $\ISS$, we suggest the following ``\textit{normative regimentation-/sanction-/reparation-based coordination toolbox}''---as a basis for  contextualized coordination mechanisms---to ensure the normatively desirable behavior in $\ISS$, i.e., to enforce norm compliant $\ISS$ behaviors. In each part, we present the intuition and briefly elaborate on how the method can be implemented in the RB--CGS settings.

\noindent {\textbf{Regimentation-Based Approach:}} The main idea behind  regimentation, as a tool for enforcing ``good'' behaviors, is to implement a mechanism that guarantees the prevention of all the norm-violating behaviors.  In RB--CGS setting, regimentation can be implemented by updating the transition system (in specific, by implementing a norm-supervisor function that updates the $d$ function) and forbidding actions that may result in a bad $\ISS$ behavior. As elaborated in \cite{dastani2017norm},  ensuring that all the bad behaviors are avoided  may result in avoiding some good behaviors as well.  Although this is a ``tight'' form of coordination, it is applicable for engineering critical $\ISS$ (e.g., for implementing industrial symbiosis on nuclear wastes, where no degree of violation is acceptable). But in a more flexible $\ISS$---and considering the sensitivity of  industrial domain to restrictive administration methods---we suggest coordination mechanisms of a more ``relax'' nature: sanction-/reparation-based methods.

\noindent {\textbf{Sanction-Based Approach:}} In sanction-based methods, the idea is to enforce good behaviors by making bad behaviors more costly\footnote{In general,  norm-aware sanctioning  refers to the imposition  of penalties/taxes when the behavior is undesirable. However, it can be also  employed for introducing negative sanctions, in terms of rewards/subsidies, to encourage desirable behaviors \cite{DBLP:conf/dagstuhl/NoriegaCFCS13}.}. I.e., bad behaviors may happen but the aim is to nudge the behavior towards  good one. In RB--CGS representation of $\ISS$, sanctions can be implemented by updating  the cost function $c$ (e.g., by increasing the accumulated cost of actions, included in a norm-violating behavior). An alternative would be to impose sanctions on firms that are able to avoid bad behaviors (i.e., sanctioning with respect to strategic potentials) or by applying preference-aware mechanisms\cite{bulling2016norm}.

\noindent {\textbf{Reparation-Based Approach:}} In general, the reparation-based approach gives the opportunity to the   firms to ``repair'' a norm-violating behavior---e.g., by postponing the sanction imposition. In $\ISS$ coordination, this is to expect that firms will return/recover the system to norm-compliant behavior as they tend to avoid being imposed with a sanction. To implement this, one may extend the set of norm-compliant $\ISS$ behaviors by including \emph{repaired} behaviors. For instance, the designer may tolerate any norm violating behavior $\lambda_v$ only if it is immediately followed by state $q_r$, where visiting $q_r$ is the result of (collectively) paying a compensation value $cv$ (lower than the sanction value $sv$). Thus, if the involved firms immediately pay $cv$ after a violation $\lambda_v$, they ensure that the behavior is repaired, hence will not be sanctioned. To fine-tune $cv$ and to determine the number of  steps that imposing a sanction can be postponed are relevant concerns for refining $\ISS$ coordination mechanisms.

%=========================%

\section{Open Problems and Engineering Challenges} \label{sec:open-Problems}

As motivated earlier, we call for scientific investigation on multiagent models for industrial symbiosis systems. Below, we elaborate on some open  problems in this new domain of study and highlight the relation to relevant multiagent techniques. 

%===============================

\noindent {\textbf{Transaction Cost Economics:}} One of the main advantages of implementing well-designed $\ISS$---in contrast to traditional  ad-hoc industrial symbiosis relations---is to reduce the total transaction cost. This can be achieved using automatized smart contracts, and in turn by eliminating  the searching and negotiation costs. In principle, such dynamics of transaction cost affect the stability and fairness of cost/benefit allocation mechanisms. We believe that this gap can be captured building on the line of contribution on \emph{cost of stability} \cite{meir2013bounding,bachrach2009cost} and by developing dynamic allocation methods, in contrast to approaches that take a static characteristic function as given.   

%===============================

\noindent {\textbf{Incentive Engineering:}} To support the implementation of an $\ISS$---with the aim to guarantee its resilience and stability---legislative bodies may introduce external support in terms of monetary incentives. This is either by  providing \emph{subsidies } to support socio-environmentally desirable $\ISS$ implementations  or by imposing \emph{taxes} to suppress undesirable  practices. To have a successful allocation of incentives, we lack policy-support tools, aware of the structure and dynamics of $\ISS$. To tackle this, we see that involved firms can be modeled as resource-controllers in an $\ISS$ game. This approach enables the applicability of incentive engineering methods for ensuring properties in boolean games  \cite{wooldridge2013incentive}.      

%===============================

\noindent {\textbf{Governance:}} In an $\ISS$, firms will get involved in commitments (in the institutional level) and can affect the state of resources (in the physical level). To be able to specify, monitor, and maintain a well-behaving $\ISS$, we need  multiagent commitment governance frameworks and interorganizational communication protocols. Following \cite{singh2013norms}, we see norms as a basis for ``\textit{administration of stakeholders by stakeholders}'' and for offering coordination as a service \cite{van2012coordination} in $\ISS$. While we elaborated on norm-based coordination approaches, specifying $\ISS$ communication languages, identifying their properties, and developing norm-based governance platforms for $\ISS$ are still open problems. Moreover, $\ISS$ are expected to be long-term and to evolve over time. Then  to capture their dynamicity  and accordingly updating the governance models, participation and feedback of industrial symbiosis  practitioners are among the key requirements. For such a purpose, integrating participatory policy analysis models \cite{mehryar2017structured} with norm-aware models \cite{tinnemeier2010programming,kaponis2006dynamic,DBLP:conf/ijcai/AlechinaDL13} would be a  suggested direction.

%===============================

%===============================

\noindent {\textbf{Organizational Characterization:}} For design and coordination of $\ISS$, a critical point is to understand, model, and reason about the dynamics of roles, relationships, and authority structures. E.g.,  in some forms of $\ISS$ the role of facilitating firms (i.e., broker agents) is crucial while in some, resource recycling plants play a key role.  In our proposed $\ISS$ model, we simply reason about action-based dynamics of $\ISS$,  thus abstract from organizational structures and notions such as roles, obligations, permissions, and prohibitions.    However, in multiplex $\ISS$ scenarios, it is more natural to introduce high-level organizational notions (1) for capturing the complex dynamics of $\ISS$ and (2) for  coordinating its behavior accordingly.  This is to move from ``\emph{ $\ISS$ as a multiagent system}'' to ``\emph{$\ISS$ as a multiagent organization}''. To this end, one may develop a mapping between multiagent organizational paradigms \cite{horling2004survey} and different forms of $\ISS$.  This calls for identifying the set of roles in various forms of $\ISS$---with respect to firms' control on resources and their strategic position---and accordingly verifying the appropriateness of organizational structures.  For instance, while hierarchical structures seem applicable for material-based $\ISS$ (due to geographical proximity of firms in an industrial cluster), a coalition-based structure might be appropriate for service-based $\ISS$ (due to involvement of potentially distant clusters). 

%===============================
\noindent {\textbf{Social Aspects, Trust, and Openness:}} While an $\ISS$ designer provides a secure platform (to nurture stable relations), the stability of relations also depends of  the dynamics of trust and  firms'  \emph{socially-bounded}  autonomy---in the sense of \cite{conte2016cognitive}. Note that in $\ISS$, we are dealing with (semi)autonomous firm managers who act with respect to (not only) the economic analysis on obtainable benefits (but also) the set of subjective preferences---that may be affected by social practices or trust dynamics \cite{dignum2015conceptual}. Thus, capturing agents' preferences and trust in  reasoning about the  behavior of $\ISS$ would be an interesting attempt for applying available trust formalizations \cite{wang2007formal,pinyol2013computational,huynh2006integrated} in a practical domain. A practical way forward would be to capture firms' subjective preferences---on the list of potential symbiotic coalitions--- and employ it as a filtering measure on  the feasibility of economically profitable instances \cite{yazdanpanah2019fisof}. Finally, while most frameworks for modeling organizations focus on a closed system, there is a lack of open multiagent models to specify $\ISS$\footnote{Formal modeling of open MAS is still an open problem. This motivates limiting the attention to a specific class, e.g., multiagent $\ISS$, and aiming to  tackle the problem under its boundaries.}. Following \cite{ostrom2015governing,dignum2007open}, we believe that  allowing \emph{the ability to} leave/join an $\ISS$ (as an industrial institution) calls for a new form of \emph{dynamic} stability/resilience analysis. For such a purpose, one may employ financial incentives \cite{kumar2000enterprise}, use social bounds \cite{huynh2006integrated}, or apply expectation-aware methods \cite{klein2003using}. 
%===============================

%====================================================
\section{Concluding Remarks}

 We presented a normative multiagent perspective on engineering ($\ISS$). This includes: (1) a formal specification that enables modeling the temporal, strategic, and resource-bounded characteristics  of $\ISS$, (2) operational semantics to develop computational models for  such systems, and (3) the sketch of mechanisms for ensuring desirable situations in $\ISS$.  We also showed the expressiveness of resource-bounded semantics for practical applications and elaborated on how RB--CGS can be combined with norm-based coordination mechanisms.  We deem that the successful application of multiagent-based methods  requires taking into account the dynamics of the application domain and fine-tuning the methods to develop fit-for-purpose technologies. Our contribution is the first attempt in proposing norm-aware resource-bounded  multiagent frameworks for specifying $\ISS$, verifying its normative behavior, and coordinating it towards desirable outcomes.  As an immediate extension, we aim to work on logical characterization  of the presented concepts and to pursue  their pilot deployment in industrial clusters.

\noindent \paragraph{\textbf{Acknowledgement:}} The first two authors acknowledge the received funding from European Union's \emph{Horizon 2020} programme under grant agreement No. 680843.
%===== END : MAIN BODY ========%

\bibliographystyle{splncs04}
\bibliography{mybibliography}

\end{document}